\documentclass{article}
\usepackage{amsfonts}
\newcommand{\beq}{\begin{equation}}
\newcommand{\eeq}{\end{equation}}
\newcommand{\dpl}{\displaystyle}
\newcommand{\sech}{\mathop{\rm sech}\nolimits}

\begin{document}

\title{Yang-Baxter maps and matrix solitons}

\date{\empty}
\maketitle

\begin{center}

{\bf V.M.Goncharenko$^{*}$, A.P.Veselov$^{**,***}$ }

\medskip

{\it $^*$Chair of Mathematics and Financial Applications,
Financial Academy, \\ Leningradsky prospect, 49, Moscow, Russia}

{\it $^{**}$Department of Mathematical Sciences, Loughborough University,
Loughborough, Leicestershire, LE 11 3TU, UK
}

{\it $^{***}$Landau Institute for Theoretical Physics, Kosygina 2,
Moscow, Russia

e-mails: mike-gonchar@mtu-net.ru,
 A.P.Veselov@lboro.ac.uk}

\end{center}


\bigskip

{\small  {\bf Abstract.} New examples of the Yang-Baxter maps (or
set-theoretical solutions to the quantum Yang-Baxter equation) on
the Grassmannians arising from the theory of the matrix KdV
equation are discussed. The Lax pairs for these maps are produced
using the relations with the inverse scattering problem for the
matrix Schr\"odinger operator.}

\bigskip

\subsection*{Introduction.}

The problem of studying the set-theoretical solutions to the
quantum Yang-Baxter equation was suggested by V.G. Drinfeld
\cite{D}. This stimulated research in this direction, mainly from
algebraic point of view (see e.g. \cite{WX},\cite{ESS}). The
dynamical aspects of this problem were discussed in the paper
\cite{V} where also a shorter term "Yang-Baxter map" for such
solutions was suggested.

In this paper we present some new examples of the Yang-Baxter maps
appeared in relation with the theory of solitons. In the case when
the solitons have the internal degrees of freedom described by
some manifold $X$ their pairwise interaction gives a map from $X
\times X$ into itself which satisfies the Yang-Baxter relation,
which means that the final result of multiparticle interaction is
independent of the order of collisions (see Kulish's paper
\cite{Kul} which is the first one we know containing such a
statement).

As an example of the equation with the soliton solutions having
non-trivial internal parameters we consider the matrix KdV
equation \beq \label{UU} U_t=3UU_x+3U_xU-U_{xxx} \eeq where $U$ is
$n\times n$ matrix. This equation was introduced in the famous P.
Lax's paper \cite{L} and was the subject of investigations in
several papers including \cite{WK, CD, G}. The related inverse
scattering problem for the matrix Schr\"odinger operator was
investigated by Martinez Alonso and Olmedilla \cite{AO1,O}.

We will show that the formulas from \cite{G} for two matrix KdV
soliton interaction can be generalized to determine some
Yang-Baxter maps on the Grassmannians $G(k,n)$ and products of two
Grassmannians $G(k,n) \times G(n-k,n).$ We produce also the Lax
pairs for these maps using the relations with the inverse
scattering problem for the matrix Schr\"odinger operator
\cite{AO1,O}.

\subsection*{Two-soliton interaction as Yang-Baxter map.}

Let us start with the definition of the Yang-Baxter map (cf.
\cite{D}, \cite{V}). Let $X$ be any set and $R$ be a map: $X
\times X \rightarrow X \times X.$ Let $R_{ij}: X^{n} \rightarrow
X^{n}, \quad X^{n} = X \times X \times .....\times X$ be the maps
which acts as $R$ on $i$-th and $j$-th factors and identically on
the others. If $P: X^2 \rightarrow X^2$ is the permutation:
$P(x,y) = (y,x)$, then $$R_{21} = P R P.$$

The map $R$ is called {\it Yang-Baxter map} if it satisfies the
Yang-Baxter relation
\begin{equation}
\label{YB} R_{12} R_{13} R_{23} = R_{23} R_{13} R_{12},
\end{equation}
considered as the equality of the maps of $X \times X \times X$
into itself. If additionally $R$ satisfies the relation
\begin{equation}
\label{U} R_{21} R = Id,
\end{equation}
we will call it {\it reversible Yang-Baxter map}.

We will actually consider the {\it parameter-dependent Yang-Baxter
maps} $R(\lambda,\mu), \lambda, \mu \in {\bf C}$ satisfying the
corresponding version of Yang-Baxter relation
\begin{equation}
\label{sYB} R_{12}(\lambda_1, \lambda_2) R_{13}(\lambda_1,
\lambda_3) R_{23} (\lambda_2, \lambda_3) = R_{23}(\lambda_2,
\lambda_3) R_{13}(\lambda_1, \lambda_3) R_{12}(\lambda_1,
\lambda_2)
\end{equation}
and reversibility condition
\begin{equation}
\label{sU} R_{21}(\mu,\lambda) R(\lambda, \mu) = Id.
\end{equation}
Although this case can be considered as a particular case of the
previous one by introducing $\tilde X = X \times {\bf C}$ and
$\tilde R (x,\lambda; y, \mu) = R (\lambda,\mu) (x,y)$ it is more
convenient for us to keep the parameter separately.

To construct the examples of the such maps consider the
two-soliton interaction in the matrix KdV equation (\ref{UU}). At
the beginning let $U$ be a general $n \times n$ complex matrix, no
symmetry conditions are assumed.

It is easy to check that the matrix KdV equation has the soliton
solution of the form $$U = 2 \lambda^2 P \sech^2 (\lambda x-
4\lambda^3 t),$$ where $P$ must be a projector: $P^2 = P.$ If we
assume that $P$ has rank 1 then $P$ should have the form
$P=\frac{\dpl \xi\otimes \eta}{\dpl (\xi,\eta)}$. Here $\xi$ is a
vector in a complex vector space $V$ of dimension $d$, $\eta$ is a
vector from the dual space $V^*$ (covector) and bracket
$(\xi,\eta)$ means the canonical pairing between $V$ and $V^*.$

To find the two-soliton solutions one can use the inverse
scattering problem for the general matrix Schr\"odinger operator
developed in \cite{AO1, O}. The corresponding formulas have been
found and analyzed in \cite{G}. In particular, it was shown that
the change of the matrix amplitudes $P$ ("polarizations") of two
solitons with the velocities $\lambda_1$ and $\lambda_2$ after
their interaction is described by the following map:
$$R(\lambda_1, \lambda_2): (\xi_1, \eta_1; \xi_2, \eta_2)
\rightarrow (\tilde{\xi_1}, \tilde{\eta_1}; \tilde{\xi_2},
\tilde{\eta_2)}$$ \beq \label{fi} \tilde{\xi_1} = \xi_1+\frac{\dpl
2\lambda_2(\xi_1,\eta_2)}{\dpl (\lambda_1-
\lambda_2)(\xi_2,\eta_2)}\xi_2, \qquad \tilde{\eta_1} =
\eta_{1}+\frac{\dpl 2\lambda_2(\xi_2,\eta_1)}{\dpl (\lambda_1-
\lambda_2)(\xi_2,\eta_2)}\eta_2, \eeq

\beq \label{fj} \tilde{\xi_2} = \xi_2+\frac{\dpl
2\lambda_1(\xi_2,\eta_1)}{\dpl (\lambda_2-
\lambda_1)(\xi_1,\eta_1)}\xi_1,\qquad \tilde{\eta_2} =
\eta_2+\frac{\dpl 2\lambda_1(\xi_1,\eta_2)}{\dpl (\lambda_2
-\lambda_1)(\xi_1,\eta_1)}\eta_1. \eeq

We claim that this map is a reversible parameter-dependent
Yang-Baxter map. This can be checked directly although the
calculations are quite long.

A better way is explained in the next section.

\subsection*{Matrix factorizations and Lax pairs.}

Suppose we have a matrix $A(x,\lambda; \zeta)$ depending on the
point $x \in X$, parameter $\lambda$ and additional parameter
$\zeta \in {\bf C},$ which we will call {\it spectral parameter.}
We assume that $A$ depends on $\zeta$ polynomially or rationally.
The case of elliptic dependence is also very interesting (see
\cite{Od}) but we will not consider it here.

Consider the product $L =  A(y,\mu; \zeta)A(x,\lambda; \zeta),$
then change the order of the factors $L \rightarrow \tilde{L} =
 A(x,\lambda; \zeta)A(y,\mu; \zeta)$ and re-factorize it as:
$\tilde{L} = A(\tilde{y},\mu; \zeta)A(\tilde{x},\lambda; \zeta) .$
Suppose that this re-factorization relation
\begin{equation}
\label{refact}
 A(x,\lambda; \zeta)A(y,\mu; \zeta) =  A(\tilde{y},\mu; \zeta)A(\tilde{x},\lambda; \zeta)
\end{equation}
uniquely determines $\tilde{x},\tilde{y}.$

It is easy to see that the map \beq \label{fact}
R(\lambda,\mu)(x,y) = (\tilde{x},\tilde{y}) \eeq determined by
(\ref{refact}) satisfies the Yang-Baxter relation. Indeed if we
consider the product $A(x_1) A(x_2) A(x_3)$ (we omit here the
parameters $\lambda_i$ and  $\zeta$ for shortness) then applying
the left hand side of (\ref{YB}) to this product we have $A(x_1)
A(x_2) A(x_3) = A(x_1^{(1)}) A(x_3^{(1)}) A(x_2^{(1)})=
A(x_3^{(2)}) A(x_1^{(2)}) A(x_2^{(2)}) = A(x_3^{(3)}) A(x_2^{(3)})
A(x_1^{(3)}).$ Similarly the right hand side corresponds to the
relations $A(x_1) A(x_2) A(x_3) = \\ A(\tilde{x}_2^{(1)})
A(\tilde{x}_1^{(1)}) A(\tilde{x}_3^{(1)})= A(\tilde{x}_2^{(2)})
A(\tilde{x}_3^{(2)}) A(\tilde{x}_1^{(2)}) = A(\tilde{x}_3^{(3)})
A(\tilde{x}_2^{(3)}) A(\tilde{x}_1^{(3)}).$ If the factorization
is unique we have $x_i^{(3)} = \tilde{x}_i^{(3)},$ which is
exactly the Yang-Baxter relation.

If a parameter-dependent Yang-Baxter map $R(\lambda,\mu)$ can be
described in such a way we will say that $A(x,\lambda; \zeta)$ is
a {\it Lax pair for} $R$. As it was shown in \cite{V} such a Lax
pair allows to produce the integrals for the dynamics of the
related transfer-maps.

Let us come back now to matrix solitons. We claim that the map
described by the formulas (\ref{fi}),(\ref{fj}) has the Lax pair
of the following form \footnote{Yuri Suris suggested a simple
explanation of this form which works also for a wide class of the
Yang-Baxter maps (see \cite{SV}).} motivated by the inverse
spectral problem for the matrix Schr\"odinger operator \cite{O}:
\begin{equation}
\label{Lax1} A(\xi, \eta, \lambda; \zeta) = I + \frac{2
\lambda}{\zeta -\lambda}\frac{\dpl \xi\otimes \eta}{\dpl
(\xi,\eta)}
\end{equation}
In the soliton theory this type of matrices were first used by
Zakharov and Shabat \cite{ZSh}.

One can check directly that re-factorization relation for this
matrix leads to the map (\ref{fi}, \ref{fj}) but we would prefer
to do this in a more general situation.

\subsection*{Generalization: Yang-Baxter maps on the Grassmannians.}

Let $V$ be an $n$-dimensional real (or complex) vector space, $P:
V \rightarrow V$ be a projector  of rank $k$: $P^2 =P$. Any such
projector is uniquely determined by its kernel $K = Ker P$ and
image $L = Im P,$ which are two subspaces of $V$ of dimensions $k$
and $n-k$ complementary to each other: $K \oplus L = V.$ The space
of all projectors $X$ of rank $k$ is an open set in the product of
two Grassmannians $G(k,n) \times G(n-k,n).$

Consider the following matrix
\begin{equation}
\label{Lax2} A(P, \lambda; \zeta) = I + \frac{2 \lambda}{\zeta
-\lambda} P
\end{equation}
and the related re-factorization relation
\begin{equation}
\label{rat}
 (I + \frac{2 \lambda_1}{\zeta -\lambda_1} P_1)(I +
\frac{2 \lambda_2}{\zeta -\lambda_2} P_2) = (I + \frac{2
\lambda_2}{\zeta -\lambda_2} \tilde{P}_2)(I + \frac{2
\lambda_1}{\zeta -\lambda_1} \tilde{P}_1)
\end{equation}
which we can rewrite in the polynomial form as
\begin{equation}
\label{ref} ((\zeta -\lambda_1) I + 2 \lambda_1 P_1)((\zeta
-\lambda_2) I + 2 \lambda_2 P_2) = ((\zeta -\lambda_2) I + 2
\lambda_2\tilde{P}_2)((\zeta -\lambda_1) I + 2 \lambda_1
\tilde{P}_1).
\end{equation}

We claim that if $\lambda_1 \neq \pm\lambda_2$ it has a unique
solution. This follows from the general theory of matrix
polynomials (see e.g. \cite{GLR}) but in this case we can see this
directly.

Indeed let us compare the kernels of both sides of the relation
(\ref{ref}) when the spectral parameter $\zeta = \lambda_1.$ In
the right hand side we obviously have $\tilde{K}_1$ while the left
hand side gives $$((\lambda_1 -\lambda_2) I + 2 \lambda_2
P_2)^{-1} K_1 = (I + \frac{2 \lambda_2}{\lambda_1
-\lambda_2}P_2)^{-1} K_1.$$

Now we use the following property of the matrix (\ref{Lax2}):
\begin{equation}
\label{property} A(P, -\lambda; \zeta) =  A(P, \lambda;
\zeta)^{-1}
\end{equation}
to have
\begin{equation}
\label{K1}\tilde{K}_1 = (I - \frac{2 \lambda_2}{\lambda_1
+\lambda_2}P_2) K_1.
\end{equation}

Similarly taking the image of both sides of (\ref{ref}) at $\zeta
= \lambda_2$ we will have
\begin{equation}
\label{L2}\tilde{L}_2 = (I + \frac{2 \lambda_1}{\lambda_2
-\lambda_1}P_1) L_2.
\end{equation}

To find $\tilde{K}_2$ and $\tilde{L}_1$ one should take first the
inverse of both sides of (\ref{rat}), use the property
(\ref{property}) and then repeat the procedure. This will lead us
to the formulas:
\begin{equation}
\label{K2}\tilde{K}_2 = (I - \frac{2 \lambda_1}{\lambda_1
+\lambda_2}P_1) K_2
\end{equation}
and
\begin{equation}
\label{L1}\tilde{L}_1 = (I + \frac{2 \lambda_2}{\lambda_1
-\lambda_2}P_2) L_1.
\end{equation}

The formulas (\ref{K1},\ref{L2},\ref{K2},\ref{L1}) determine a
parameter-dependent Yang-Baxter map on the set of projectors. One
can easily check that for $k=1$ one has the formulas (\ref{fi},
\ref{fj}) for two matrix soliton interaction.

If we supply now our vector space $V$ with the Euclidean
(Hermitian) structure and consider the self-adjoint projectors $P$
of rank $k$ then the corresponding space $X$ will coincide with
the Grassmannian $G(k,n):$ such a projector is completely
determined by its image $L$ (which is a $k$-dimensional subspace
in $V$ and thus a point in $G(k,n)$) since the kernel $K$ in this
case is the orthogonal complement to $L.$

The corresponding Yang-Baxter map $R$ on the Grassmannian is
determined by the formulas
\begin{equation}
\label{L11}\tilde{L}_1 = (I + \frac{2 \lambda_2}{\lambda_1
-\lambda_2}P_2) L_1,
\end{equation}
\begin{equation}
\label{L22}\tilde{L}_2 = (I + \frac{2 \lambda_1}{\lambda_2
-\lambda_1}P_1) L_2.
\end{equation}

It would be very interesting to investigate the dynamics of the
corresponding transfer-maps \cite{V}. As we have shown here the
Lax pair for them is given by (\ref{Lax2}).

\section*{ Acknowledgements}

The second author (A.P.V.) is grateful to the organizers and
participants of the NEEDS conference and NATO Advanced Research
Workshop on "New Trends in Integrability and Partial Solvability"
in Cadiz (10-15 June 2002) and SIDE-V conference in Giens (21-26
June 2002) where these results were first presented and especially
to P.P. Kulish, A.B. Shabat and Yu. Suris for stimulating and
helpful discussions.

\end{document}